\documentstyle[twocolumn,prb,aps,epsf]{revtex}

\begin{document}
\draft

\twocolumn[\hsize\textwidth\columnwidth\hsize\csname @twocolumnfalse\endcsname

\title{Comparison of different ladder models}
\author{E.~Jeckelmann$^*$, D.J.~Scalapino$^\dagger$, S.R.~White$^*$}
\address{$^*$Department of Physics, University of California, Irvine, CA
92697 \\
$^\dagger$Department of Physics,
University of California,
Santa Barbara, CA 93106
}
\maketitle
\begin{abstract}
Using density matrix renormalization group calculations,
we compare results obtained for the $t-J$, one-band Hubbard and 
three-band Hubbard models of a two-leg $CuO$ ladder.
Spin and charge gaps, pair binding energies, and effective pair
hoppings are calculated for a wide range of parameters.
All three models have an insulating state at a filling
corresponding to one hole per $Cu$ site.
For physically relevant parameters their spin gaps are similar
in size but they exhibit quite different charge gaps.
We find that the binding energy of a pair
of doped holes is of the order of the undoped ladder spin gap
for all three models.
The main difference between the models is the size of
the effective pair hopping, which is significantly
larger in the three-band model 
with parameters appropriate for $CuO$ materials
than in the other two models.
\end{abstract}


]



The 2-leg $CuO$ ladder materials have provided an interesting testing
ground for ideas originally formulated to describe the 2D cuprates.
\cite{DR96} In particular, the undoped, 2-leg ladder material
$SrCu_2O_3$  exhibits a spin gap \cite{Tak92}  
and the doped $(SrCa)_{14} Cu_{24}O_{41}$ ladder
material \cite{Ueh96} can become superconducting under high pressure.
The reduced dimensionality of this system has allowed for detailed
numerical studies of both the $t-J$ and the one-band Hubbard models of a
2-leg ladder. In fact, it was a numerical Lanczos study \cite{DRS92} of a
2-leg
$t-J$ ladder which first suggested that the doped system might exhibit
superconducting pairing. Since then, analytic calculations 
\cite{GRS94,BF96} as well as density matrix renormalization group
(DMRG) calculations \cite{NSW96} on long one-band Hubbard ladders have
shown the $d_{x^2-y^2}$-like structure of the pairs and their power law
correlations.  Here we extend the DMRG approach to study a three-band Hubbard
model of a $CuO$ 2-leg ladder with the goal of comparing the $t-J$,
one-band, and three-band Hubbard models. 

The three-band Hubbard model we have studied has a hole Hamiltonian which
can formally be thought of in terms of the ``3-leg'' ladder shown in 
Fig.~1. Alternating sites on the top and bottom legs have
$Cu(d_{x^2-y^2})$ and $O(p_x)$ orbitals, while the center leg has $O(p_y)$
orbitals on the sites bridging the $Cu(d_{x^2-y^2})$ orbitals.
With the orbital phase convention we have chosen, all the $O-Cu$
hopping matrix elements are $-t_{pd}$. 
The difference in the hole site energies is $\Delta =
\varepsilon_p-\varepsilon_d$, and there is an onsite $Cu$
Coulomb interaction $U_d$. The undoped system has one hole per $Cu$. In this
state, if $U_d$ is large compared to $\Delta$, the charge gap is set by
$\Delta$ and the system is said to be a charge gap insulator.~\cite{ZSA85}
Various parameter values have been suggested,~\cite{HSSJ,MAM90,Mar96,And}
with typical ones having 
$\Delta/t_{pd}=$2 to 3, $U_d/t_{pd}=8$, and $t_{pd}$ ranging from 1.3eV
to 1.8 eV.
Throughout this paper we will confine ourselves to
an isotropic ladder and leave the case in which the rung parameters differ
from the leg parameters to a future study.~\cite{Bul96} 

The 2-leg Hubbard model has a hopping $-t$ between the near neighbor ladder
sites (along the legs and across the rungs) and an onsite Coulomb
interaction $U$. Here the undoped state corresponds to half-filling. At
half-filling, the charge gap depends upon $U$ and is said to be a
Mott-Hubbard gap.\cite{ZSA85} 
Typical parameters \cite{HSSJ,MAM90,Mar96,And}
are $U/t=10$ to $12$ and $t=0.4$ to $0.45$eV.

The $t-J$ model has a near neighbor hopping term $-t$ with a restriction
that no site can have two fermions. The exchange interaction in the $t-J$
model has the usual form
\begin{equation} 
J\sum_{\langle ij\rangle}\ \left(\vec S_i \cdot \vec S_j - \frac{1}{4}\ n_i
n_j\right)
\label{one}
\end{equation}
with $\vec S_i = \frac{1}{2}\ c^\dagger_{is} \sigma_{ss^\prime}
c_{is^\prime}$ and $n_i=c^\dagger_{i\uparrow} c_{i\uparrow} +
c^\dagger_{i\downarrow} c_{i\downarrow}$. At half-filling, the $t-J$ model
is just the Heisenberg 2-leg ladder and has an infinite charge gap. Typical
parameters \cite{HSSJ,MAM90,Mar96,And} are $J/t=0.3$ and $t=$0.4 to 0.5eV.

We use DMRG techniques \cite{White92} to study long 
ladders (up to $64\times2$ sites in the $t-J$ and one-band Hubbard 
models and up to $16\times2$ $Cu$
sites in the three-band Hubbard model) with open boundary
conditions. DMRG has been shown to be a very accurate method
to study ladder 
systems.\cite{NSW96,Bul96,NWS94,WNS94} 
Here we use up to 1200 states per block

\begin{figure}[h]
\epsfxsize=2.375 in\centerline{\epsffile{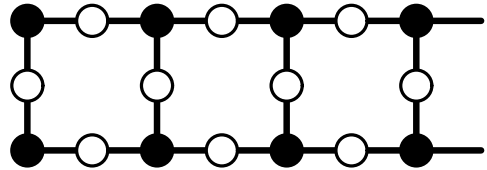}}
\caption{Schematic of a $CuO$ ladder.
Here the solid circles represent  $Cu(d_{x^2-y^2})$
orbitals and the open circles represent $O (p_x)$ orbitals along
the upper and lower legs and $O (p_y)$ orbitals
on the rungs. There is a hopping matrix element $-t_{pd}$ between
the $O$ and $Cu$ sites as shown by the solid lines.
The energy difference between the $O$
and $Cu$ sites is $\Delta=\varepsilon_p-\varepsilon_d$ and there
is an onsite $Cu$ Coulomb interaction $U_d$.}
\label{fig1}
\end{figure}

\noindent and extrapolate DMRG results for energies to extract the limit
of zero truncation error and check the precision of our
calculations.\cite{Bonca} 
In addition, a few tests performed on small clusters with periodic 
boundary conditions show a perfect agreement with exact diagonalization 
results.\cite{OS88}


We begin with the undoped ladder and calculate the spin gap
\begin{equation}
\Delta_s = E_0\left(S_z=1\right) - E_0 \left(S_z=0\right)
\label{two}
\end{equation}
Here $\Delta_s$ is the difference in energies between the spin 1 and spin 0
ground states. For the Heisenberg 2-leg ladder \cite{WNS94}
\begin{equation}
\Delta_s \simeq \frac{J}{2}
\label{three}
\end{equation}
which is shown as the solid curve in Fig.~2, where we have plotted
$\Delta_s$ versus $4t/J$. The dashed curve in Fig.~2 shows results for a
$32\times 2$, half-filled Hubbard ladder. For large values of $U/t$, the
Hubbard model at half-filling maps to the Heisenberg model with $J=4t^2/U$.
Thus in Fig.2, the two curves approach each other at large values of $U/t$.
Using twice the spin gap as a measure of the effective exchange
interaction, we see from Fig.~2 that the strong coupling expression
$J=4t^2/U$ for the one-band Hubbard model
 overestimates the strength of the exchange interaction for
physically relevant values of the parameters. However, we can use
$\Delta_s$ as a unit of energy in comparing the $t-J$ and one-band Hubbard
models, each in their relevant physical parameter regimes. 
Thus, for an infinite 2-leg ladder,
the $t-J$ model with $J/t=0.3$ has $\Delta_s/t \cong 0.15$
while the one-band Hubbard model with $U/t=12$ has $\Delta_s/t \cong 0.12$
as seen in Table~1. 
Taking 
$t=0.45$eV, twice these spin gap energies give reasonable
effective exchange couplings of 0.135eV (1600K) and 0.11eV (1300K) for the
$t-J$ and one-band Hubbard ladders, respectively. Naturally, either of
these two could be further adjusted by using a different value of $t$, 
but our point is simply that they are in the correct range.
We also note that the spin gap $\Delta_s/t \cong 0.11$ 
of the one-band Hubbard model with $U/t=6$ is similar to the value 
obtained for $U/t=12$ (see Table~1). 

\begin{figure}[h]
\epsfxsize=3.375 in\epsfysize=2 in\centerline{\epsffile{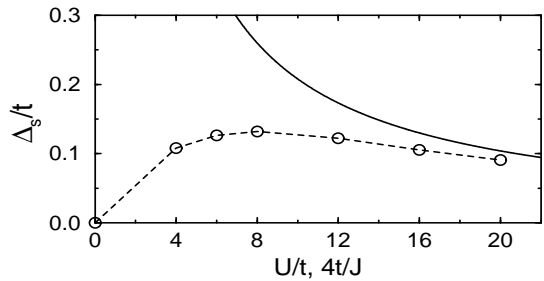}}
\caption{The spin gap $\Delta_s/t$ versus $4t/J$ for
the $t-J$ (Heisenberg) model (solid line)
and $U/t$ for the one-band Hubbard model (dashed line) at
half-filling on a $32\times2$ ladder.}
\label{fig2}
\end{figure}

\begin{figure}[h]
\epsfxsize=3.375 in\centerline{\epsffile{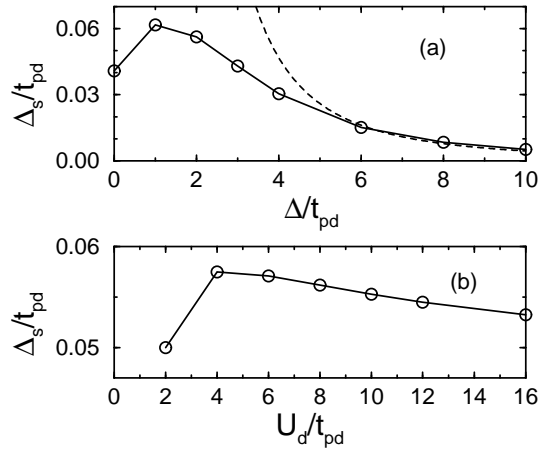}}
\caption{(a) The spin gap $\Delta_s/t_{pd}$ versus
$\Delta/t_{pd}$
for the 3-band Hubbard model at a filling of one hole per $Cu$, with
$U_d/t_{pd}=8$. The dashed line is the strong coupling limit,
Eq.~(\ref{four}).
(b) The spin gap energy $\Delta_s$ versus $U_d/t_{pd}$ with
$\Delta/t_{pd} = 2$ for the 3-band Hubbard model at a filling of one hole
per
$Cu$. All results are for an $8\times2$ $Cu$ ladder.}
\label{fig3}
\end{figure}

Results for the spin gap $\Delta_s$ of the three-band Hubbard model at a
filling of one hole per $Cu$ are shown in Figs 3(a) and (b). In Fig.~3(a),
$\Delta_s/t_{pd}$ is plotted versus $\Delta/t_{pd}$ for a ladder containing
$8\times 2\ Cu$ sites with $U_d/t_{pd}=8$. 
In strong coupling, the $Cu-Cu$ exchange
interaction has the form
\begin{equation}
J_{cu} = 4\left(\frac{t^2_{pd}}{\Delta}\right)^2 \left[\frac{1}{U_d} +
\frac{1}{\Delta}\right]
\label{four}
\end{equation}
The dashed curve in Fig.~3(a)
shows $J_{cu}/2$ and one sees that the spin gap of the three-band
model approaches the strong coupling $J_{cu}/2$ result  at large values of the
$\Delta=\varepsilon_p-\varepsilon_d$ splitting.
However, just as 
for the one-band Hubbard model, the strong-coupling expression for the
exchange interaction substantially overestimates it in the physically
relevant region of parameters. The dependence of $\Delta_s$ on $U_d/t_{pd}$
for $\Delta/t_{pd}=2$ is shown in Fig.~3(b). At
large values of $U_d/t_{pd}$, the exchange is dominated by the
$\Delta$ term in Eq.~(\ref{four}) as
expected for a charge transfer insulator. 
Taking $\Delta/t_{pd} = 2$ and
$U_d/t_{pd} =8$ we find that $\Delta_s/t_{pd}$ extrapolates 
to $\cong 0.035$ for an infinite $CuO$ ladder (see Table~1),
which for $t_{pd}=1.5$eV gives
$J_{cu} \equiv 2\Delta_s \cong 0.11$eV (1300K). 
Thus, even though the region of relevant
physical parameters does not lend itself to a simple strong coupling
expansion which relates the three models, the spin gaps for the three
models are in fact similar in size and provide a useful energy
scale describing the strength of the exchange coupling. 

The charge gaps for the insulating (undoped) phases, 
however, are quite different.
The insulating state of the $t-J$ model has an infinite charge gap
reflecting the constraint that no site can have two fermions. 
In the Hubbard models, the charge gap is defined by
\begin{equation}
\Delta_{c} = \left(E_0(2) + E_0(-2) - 2E_0(0)\right)/2,
\label{five}
\end{equation}
where $E_0(n)$ is the ground state energy of a ladder
with $n$ holes relative to the undoped ladder. 
The charge
gaps for the half-filled, one-band Hubbard model and the three-band Hubbard
model with one hole per $Cu$ are plotted versus the onsite Coulomb 
interaction in Fig.~4(a). At large
values of $U$, the charge gap in the one-band Hubbard model varies as $U$,
while as expected, the charge gap of the three-band Hubbard model saturates
at a value set by $\Delta$ when $U_d$ becomes
very large. 
As noted above, the spin gap energy provides a useful unit of energy in 
comparing the different models. In Table~1 we show the charge gap in units
of the spin gap in the one-band Hubbard model for $U/t=$ 6 and 12 and in
the three-band Hubbard model for the physical parameters $\Delta/t_{pd}=$
2 and 3 with $U_d/t_{pd}=8$. 
One can use either $U/t=$ 6 or 12 to
reproduce the three-band model $\Delta_c/\Delta_s$ ratio
depending on the value of $\Delta/t_{pd}$.  
Taking the same values of $t$ and $t_{pd}$ as above, we obtain
$\Delta_c=$ 1.4 to 3.9~eV.
The experimental value of the charge gap is still debated\cite{Mar96}
but is of the same order of magnitude (2-5eV).

\begin{figure}[h]
\epsfxsize=3.375 in\centerline{\epsffile{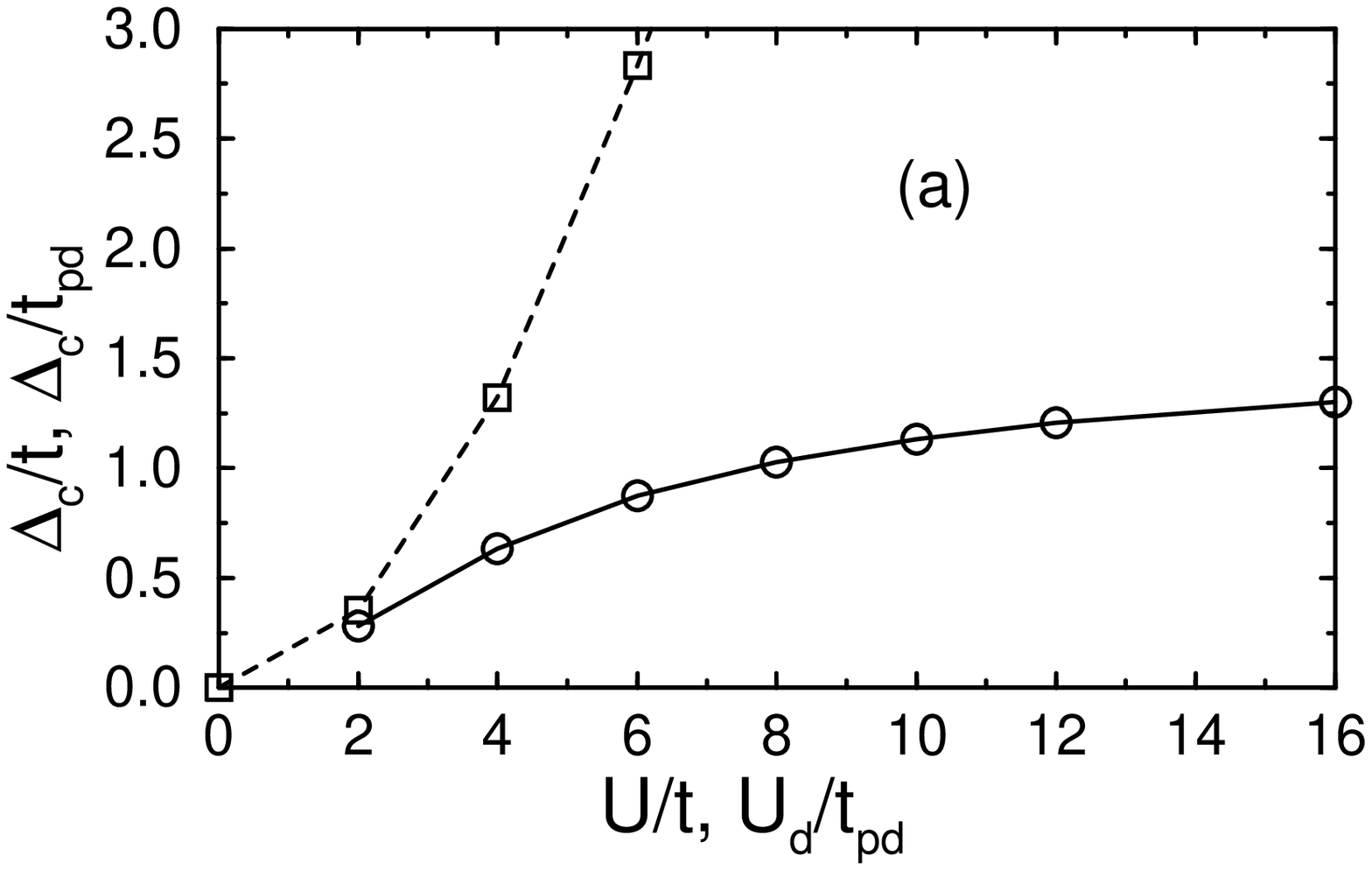}}
\epsfxsize=3.375 in\centerline{\epsffile{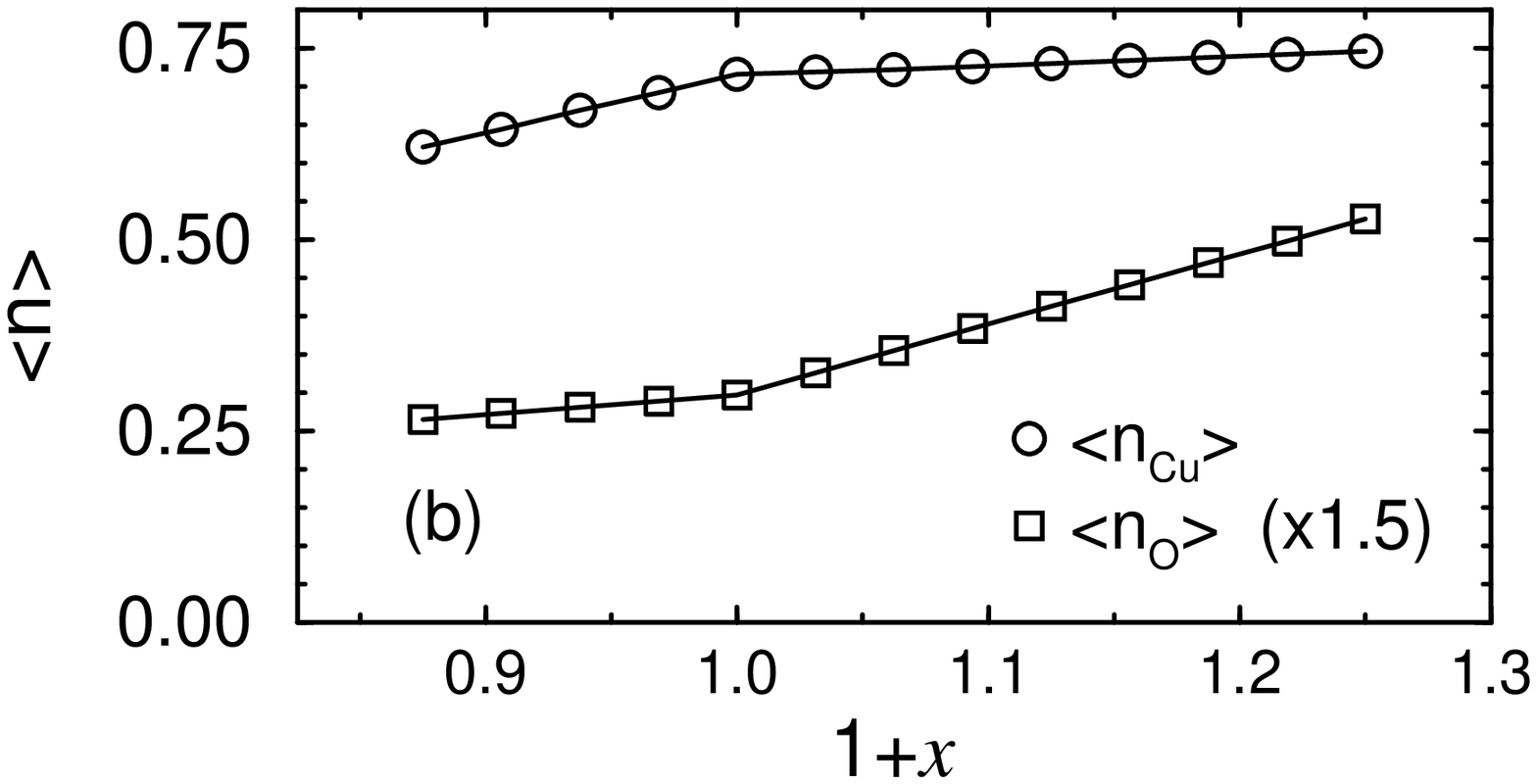}}
\caption{(a) The charge gap $\Delta_c$ versus $U/t$ for
the one-band (dashed) Hubbard model on a $32\times2$ ladder
and versus $U_d/t_{pd}$ for the
three-band (solid) Hubbard model with $\Delta/t_{pd}=2$
on a ladder containing $8\times2$ $Cu$ sites.
Note that
$\Delta_c$ is measured relative to $t$ for the one-band Hubbard
model and relative to $t_{pd}$ for the three-band Hubbard model. (b)
The hole occupation per $Cu\ \langle
n_{Cu}\rangle$ and $O\ \langle n_0\rangle$ versus the hole concentration
$1+x$ per $CuO_{1.5}$ unit cell on a ladder containing
$16\times2$ $Cu$ sites with $U_d/t_{pd}=8$ and $\Delta/t_{pd}=2$.}
\label{fig4}
\end{figure}

The ``charge transfer gap'' behavior of
the three-band model is also seen in Fig.~4(b) where we have plotted the hole
occupation per $Cu$ and per 1.5 $O$ versus the hole concentration $1+x$ per
$CuO_{1.5}$ unit cell for the physical parameters
$U_d/t_{pd}=8$ and $\Delta/t_{pd}=2$. For $x <0$,
the holes go primarily onto the $Cu$ sites (72\% of the hole
density is on these sites) while for $x>0$, the $O$ sites
are favored with 88\% of the additional hole density
going to these sites. For example, for $x=0.25$, the average
hole density on the $Cu$ site
is increased by only 0.03 (4\%) while that on an oxygen site
is increased by about 0.15 (77\%), or 0.22 per $O_{1.5}$,
from the result for the undoped ladder ($x=0$). 
This behavior can easily be understood. For $x\leq 0$,
only $Cu$ orbitals are occupied because they have lower energy than
the $O$ orbitals. 
(The hybridization of the $Cu$ and $O$ orbitals due to
the finite hopping $t_{pd}$ is responsible for the fractional
density on $Cu$ and $O$ sites.)
For $x>0$, all $Cu$ orbitals are occupied
by at least one hole. The energy to put a second hole on
one of these orbitals is set by $U_d$, while the energy to put
a hole on the $O$ orbitals is set by $\Delta$.
Thus, for the parameters considered here ($\Delta < U_d$), 
additional holes go onto $O$ orbitals.
In the opposite limit $U_d < \Delta$, we have found
that holes go primarily on $Cu$ orbitals for $x>0$,
as expected.

It is interesting to note that magnons also go primarily on $Cu$ sites 
(76\% of the spin density is on these sites). This distribution
is very close to the hole distribution of the undoped ladder and
remains constant for all doping studied $-0.125 \leq x \leq 0.25$. 
This result suggests that low-energy spin excitations involve only 
unpaired holes in (hybridized) $Cu$ orbitals even when $x\neq0$. 
According to Zhang and Rice \cite{ZR88},
each doped hole is
locked in a singlet state with another holes in a $Cu$ orbital and
does not contribute to spin excitations at low energy. 
Our results certainly support this scenario although it is not obvious
that their argument based on a $CuO_4$ cluster is valid for 
the lattice configuration shown in Fig.~1.


Next we consider the two-hole pair binding, defined by
\begin{equation}
\Delta_{pb} = E_0(2) + E_0(0) - 2E_0(1)
\label{six}
\end{equation} 
if the quantity on the right-hand side is positive and
$\Delta_{pb} = 0$ otherwise.
It should be noted that the dependence of the pair binding energy
on system size is significant.
Moreover, $\Delta_{pb}$ generally increases when the ladder length
increases, while other quantities, such as the spin gap, decrease.
Therefore, when comparing values of $\Delta_{pb}$ one should always
keep the corresponding system size in mind.
In Fig.~5(a) $\Delta_{pb}$ is plotted 
versus $4t/J$ for the $t-J$ model and $U/t$ for the one-band Hubbard model
on a $32\times2$ ladder.
Just as previously found for
the spin gap, the pair binding energy for the $t-J$ model approaches that
of the one-band Hubbard model at large values of $U/t$. Furthermore,
although the two models have very different charge gaps, the scale of their
pair binding energies in the physically relevant parameter region is set by
$\Delta_s$.
This is clearly illustrated in Fig.~5(b), where we

\begin{figure}[h]
\epsfxsize=3.375 in\centerline{\epsffile{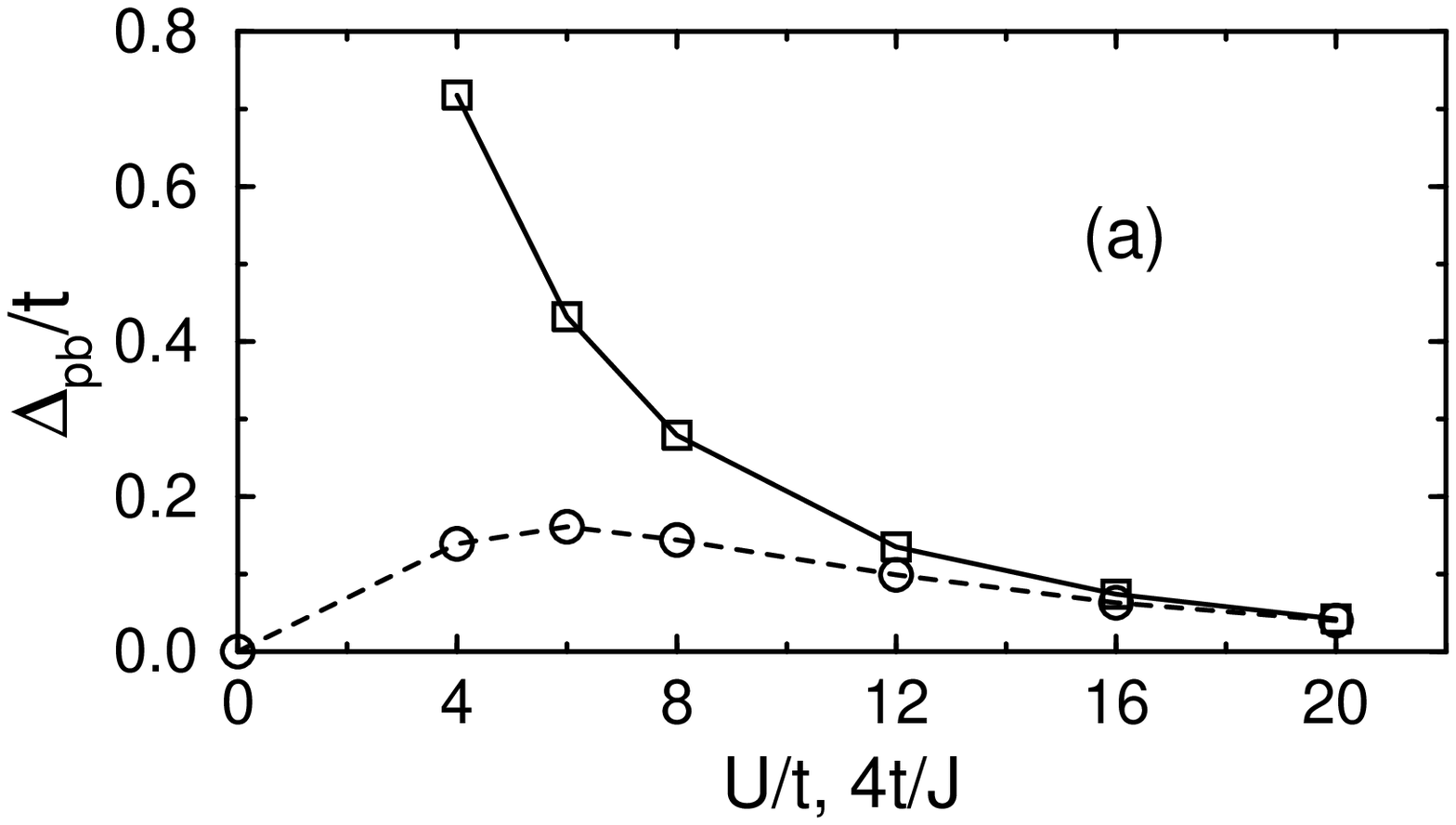}}
\epsfxsize=3.375 in\centerline{\epsffile{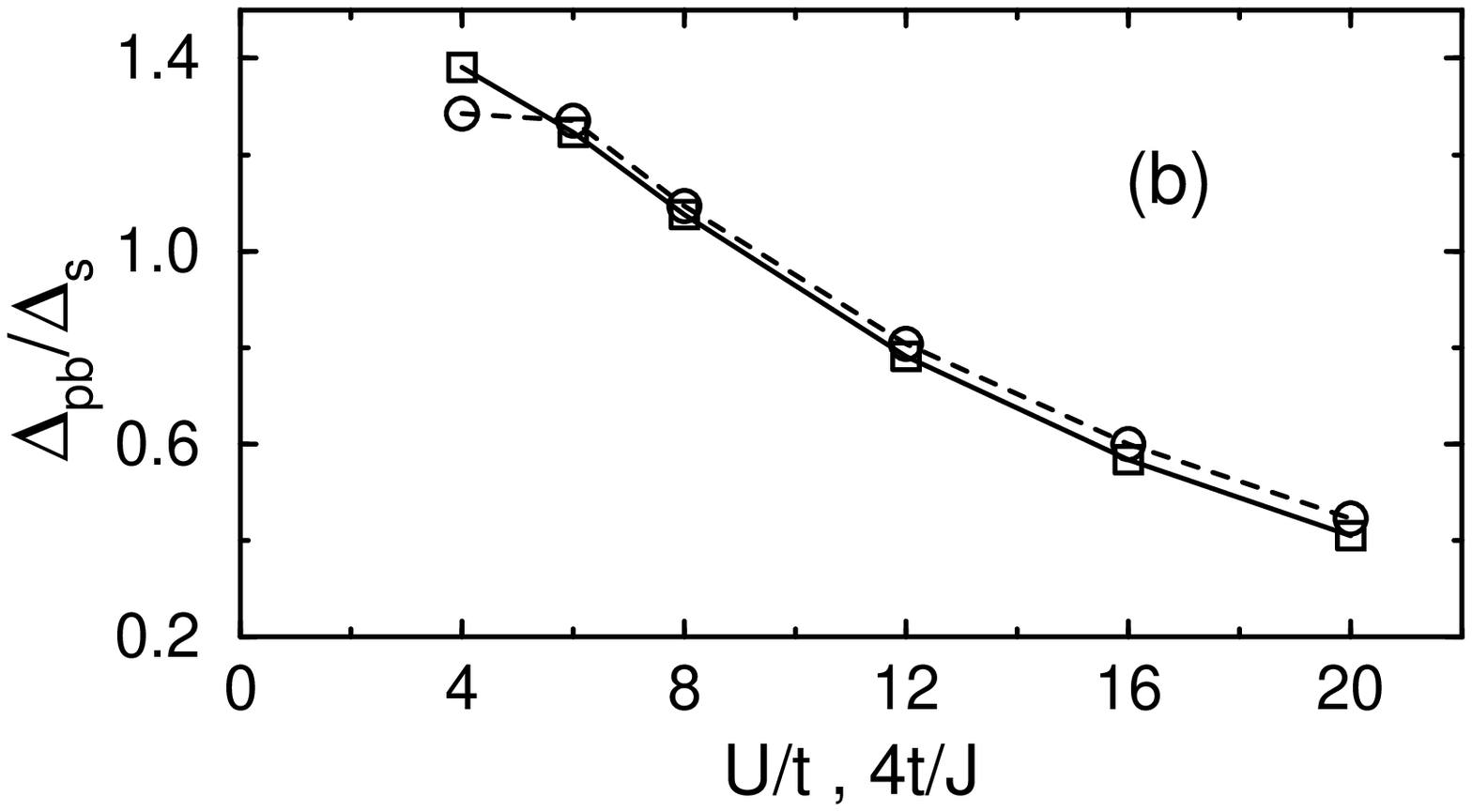}}
\caption{(a) The hole pair binding energy $\Delta_{pb}/t$
versus $4t/J$ for the $t-J$ model (solid) and $U/t$ for the
one-band Hubbard model (dashed) on a $32\times2$ ladder.
(b) The ratio $\Delta_{pb}/\Delta_s$ in both models.}
\label{fig5}
\end{figure}

\noindent show the pair binding energy
in units of the spin gap as a function of $U/t$ in the one-band Hubbard
model and as a function of $4t/J$ in the $t-J$ model on a $32\times2$ ladder.
We note that the ratio $\Delta_{pb}/\Delta_s$ is surprisingly similar in both
models despite the widely different behavior of $\Delta_{pb}$ and
$\Delta_s$ shown in Figs.~2 and 5(a).
As previously
discussed,\cite{NSW96,NWS94} the pair wave function for both of these models
has a $d_{x^2-y^2}$-like form.
Moreover, we have found that the structure of single holes and hole pairs
in the one-band Hubbard model with $U/t=8$ is identical to those observed 
in the $t-J$ model.\cite{WS97}

Results for the pair binding energy of the three-band Hubbard model are
shown in Figs 6(a) and (b). For $U_d/t_{pd} =8$, the pair binding peaks for
$\Delta/t_{pd} \cong 2$ and as shown in Fig.~6(b),
the pair binding energy increases as $U_d/t_{pd}$ increases. Similar results
for a $Cu_4O_8$ cluster were found from Lanczos calculations.\cite{OS88,Hir89}
More recently, Martin \cite{Mar96a} has discussed pairing on small
clusters in terms of rehybridization associated with the charge-transfer
channels.
A recent Quantum Monte Carlo study of the three-band Hubbard model\cite{GGZ97}
yields similar results for clusters with up to $6\times6$ $Cu$ sites.
For the physical range of parameters appropriate to the cuprates, 
one also has $\Delta_{pb}\sim \Delta_s$ in the three-band model. 
However, the ratio $\Delta_{pb}/\Delta_s$ extrapolated for a ladder 
of infinite length is clearly larger than in both other models with 
their typical parameters, as shown in Table 1. 
A better agreement with the three-band Hubbard model results
is obtained by taking $U/t=6$ for the one-band Hubbard model. 
In addition, in the three-band Hubbard model
the hole pair wave function has a $d_{x^2-y^2}$-like form as
determined from measurements of the rung-rung and rung-leg pair field
correlations. 

We have also measured the pair binding energy versus the hole concentration
$x$ relative to the undoped ladder. 
In this case the pair binding energy is defined by
\begin{eqnarray}
\Delta_{pb}(x) = E_0(2n) + E_0(2n-2) - 2E_0(2n-1), \,\,  n>0
\label{sixb}
\eqnum{6'}
\end{eqnarray} 
\begin{eqnarray}
\Delta_{pb}(x) = E_0(2n) + E_0(2n+2) - 2E_0(2n+1), \,\, n < 0
\eqnum{6''}
\label{sixc}
\end{eqnarray} 
Here $x$ is equal to $2n$ divided by the number of holes in 
the undoped ladder. 
Results for the pair binding energy $\Delta_{pb}$ relative to
the undoped spin gap $\Delta_s$ are shown in Fig.~7(a) for the three
models. 
There is no striking difference between the three models for $x>0$.
We note, however,  that while the Hubbard model is particle-hole symmetric
about half-filling, one can only add holes to the $t-J$ model and we find
no evidence for pair binding for $x<0$ in the 3-band Hubbard model with
$U_d/t_{pd}=8$ and $\Delta/t_{pd}=2$. 
We are not sure of the exact nature of the ground state  
of the three-band Hubbard model with less than one hole per $Cu$ site.
In fact, we have not been able to investigate this regime as thoroughly 
as the $x\geq0$ regime because density matrix
renormalization group calculations are much harder and less accurate
in this case.
However, our numerical simulations
strongly suggest that the $x<0$ and $x>0$ regimes are quite different.
As noted previously, holes go primarily on $Cu$ orbitals
for $x<0$ while doped holes are on $O$ orbitals for $x>0$ 
(see Fig.~4(b)).
Thus, it is possible that the effective interactions 
between doped particles are different for both regimes.

\begin{figure}[h]
\epsfxsize=3.375 in\centerline{\epsffile{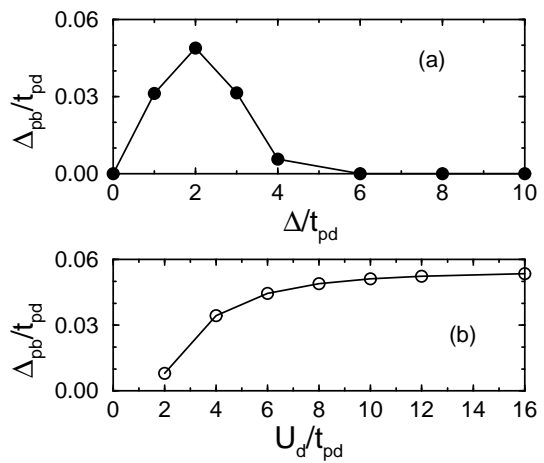}}
\caption{(a) The pair binding energy $\Delta_{pb}$
versus $\Delta/t_{pd}$ with
$U_d/t_{pd}=8$ for the 3-band Hubbard model. (b) The pair binding energy
versus $U_d/t_{pd}$ with
$\Delta/t_{pd} = 2$ for the 3-band Hubbard model.
These calculations are for an $8\times2$ $Cu$ ladder.}
\label{fig6}
\end{figure}

 \twocolumn[\hsize\textwidth\columnwidth\hsize\csname
@twocolumnfalse\endcsname

\begin{table}
\caption{Spin gap $\Delta_s$ (in units of the bare
hopping term $t$ or $t_{pd}$),
and charge gap $\Delta_c$, pair binding energy $\Delta_{pb}$, effective
pair hopping $t_{eff}$ and effective magnon hopping $v_{eff}$
(in units of the spin gap) obtained
by extrapolating to a ladder of infinite length.}
\begin{tabular}{ccccccc}
Model & Parameters & $\Delta_s$ & $\Delta_c/\Delta_s$ &
$\Delta_{pb}/\Delta_s$ & $t_{eff}/\Delta_s$ & $v_{eff}/\Delta_s$ \\ \hline
$t-J$ & $J/t=0.3$ & 0.151 & $\infty$ & 0.71 & 2.4 & 5.0\\ \hline
1-band & $U/t=12$ & 0.116 & 70 & 0.83 & 5.4 & 8.5\\ \hline
1-band & $U/t=6$ & 0.111 & 25 & 1.4 & 12 & 24 \\ \hline
3-band & $\Delta/t_{pd}=3$, $U_d/t_{pd}=8$ & 0.030 & 58 & 1.2 & 12 & 7.5 \\
\hline
3-band & $\Delta/t_{pd}=2$, $U_d/t_{pd}=8$ & 0.035 & 29 & 1.6 & 18 &
11\\
\end{tabular}
\end{table}

 ]

The decrease in the pair binding energy with $x$
correlates with the decrease in the near-neighbor spin-spin
correlations $\langle\vec S_i \cdot \vec S_j\rangle$ shown in Fig.~7(b).
(Note that for the three-band Hubbard model we show the 
spin-spin correlations between near-neighbor $Cu$ sites.)
Thus, the decrease in the pair binding energy with increased hole
concentration reflects the destruction of the underlying exchange
correlations by the added holes.
It should be noted that the differences between the values
of $\langle\vec S_i \cdot \vec S_j\rangle$ in the three models
are mostly due to the increase of charge fluctuations
in going from the $t-J$ model to the one-band Hubbard model
and then to the three-band Hubbard model. 
These charge fluctuations reduce the local 

\begin{figure}[h]
\epsfxsize=3.375 in\centerline{\epsffile{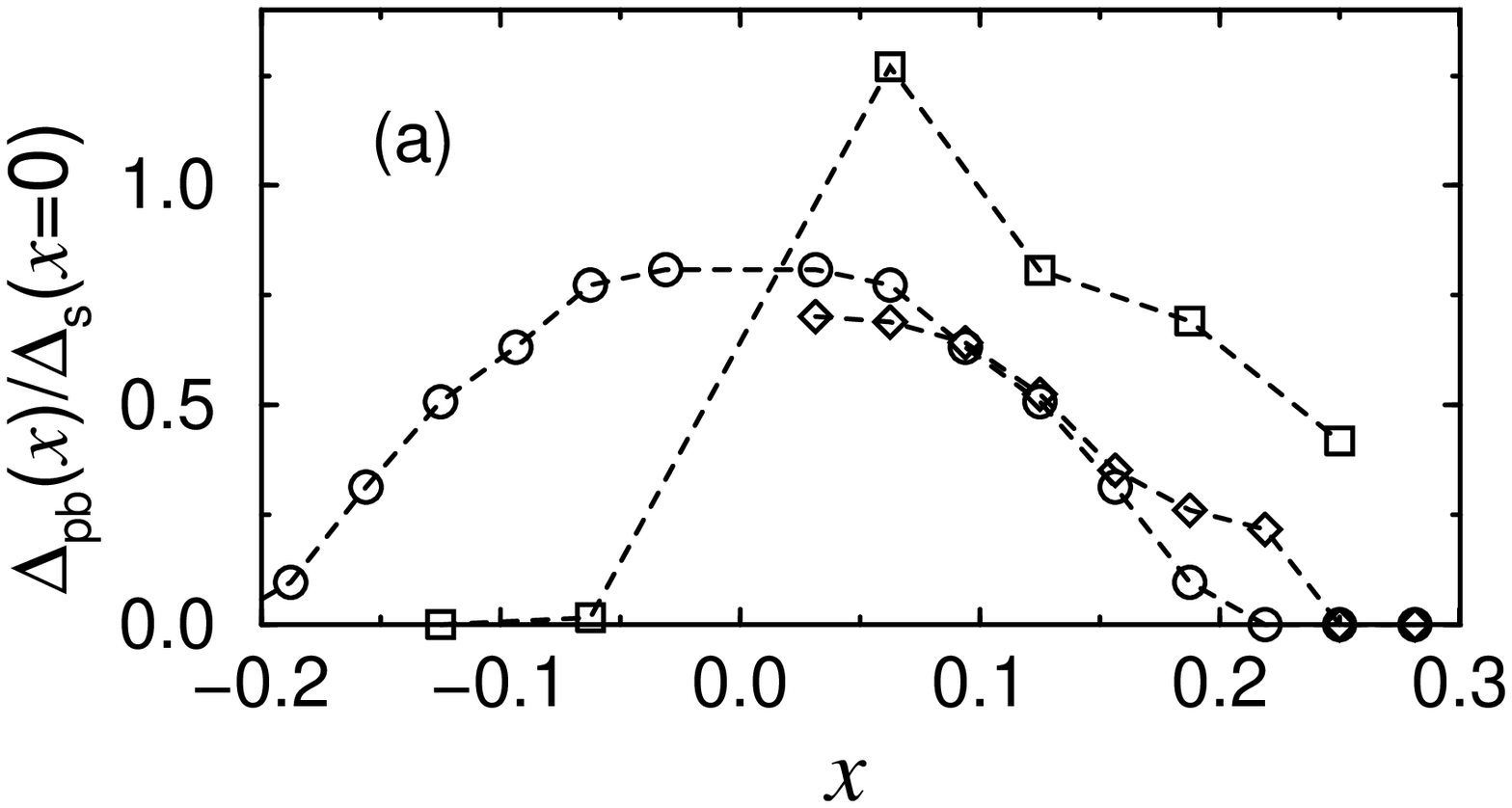}}
\epsfxsize=3.375 in\centerline{\epsffile{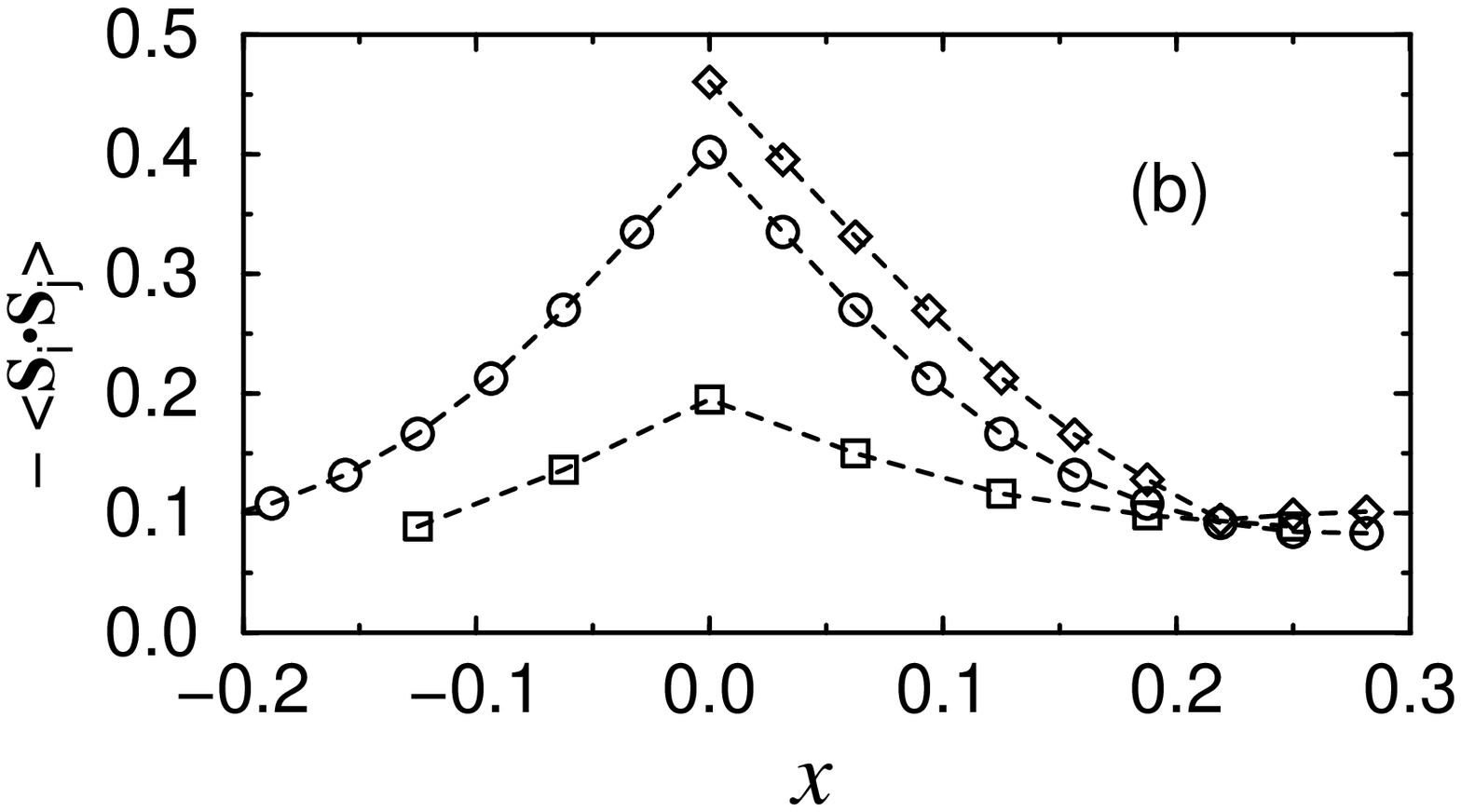}}
\caption{(a) The ratio of the pair binding energy to
the undoped spin gap versus hole doping.
The diamonds are for a $32\times2$ $t-J$ ladder with $J/t=0.3$.
The circles are for a one-band $32\times2$ Hubbard ladder with $U/t=12$.
The squares are for the three-band Hubbard model with $U_d/t_{pd}=8$ and
$\Delta/t_{pd}=2$ on a $16\times 2$ $Cu$ ladder.
(b) The near-neighbor $Cu$ spin-spin
correlation function $\langle\vec S_i\cdot \vec S_j\rangle$ versus $x$ for
the three models.}
\label{fig7}
\end{figure}

\noindent magnetic moments 
$\langle\vec S_i^2\rangle$ and thus the absolute value
of $\langle\vec S_i \cdot \vec S_j\rangle$.
However, 
as noted previously, the effective exchange coupling 
between spins is similar in the three models.

We have also calculated the effective hopping parameter $t_{eff}$
of a hole pair from the dependence of the the pair energy obtained 
in ladders of different lengths.
We define the pair energy by
\begin{equation}
\varepsilon_p = E_0(2) - E_0(0).
\label{seven}
\end{equation}
In ladders with open boundary conditions,
the energy of the pair varies as
\begin{equation}
\varepsilon_p(L_{eff}) = \varepsilon_p(\infty)+ t_{eff}\left(\frac{\pi}
{L_{eff}+1}
\right)^2
\label{eight}
\end{equation}
where the effective system length $L_{eff}$ differs from the actual ladder 
length
$L$ because of end effects.
For a given system, 
$L_{eff}$ can be determined from the wavelength $\lambda=2(L_{eff}+1)$ of 
the charge density distribution of the pair.
Fig.~8 shows a plot of the ground state energy of a hole pair versus 
$(L_{eff}+1)^{-2}$ for the three different models.
$t_{eff}$ is equal to the slope of the lines in Fig.~8 divided by
$\pi^2$. 
We have found that the difference $\delta L = L_{eff}-L$ tends to a constant 
for large systems. Thus, in practice we can substitute $L+\delta L$ for
$L_{eff}$ in Eq.~\ref{eight} and use $\delta L$ as a fit parameter.

\begin{figure}[h]
\epsfxsize=3.375 in\centerline{\epsffile{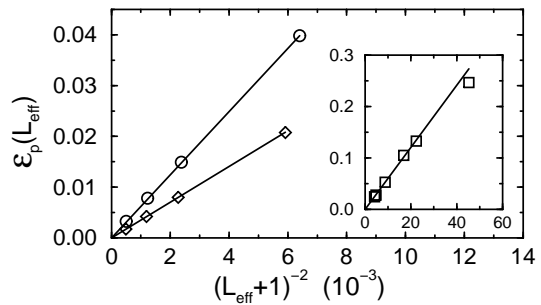}}
\caption{Plot of the hole pair energy versus
$(L_{eff}+1)^{-2}$ for the three different models.
The notation is the same as in Fig.~\ref{fig7}.
Actual ladder lengths $L$ range from 5 to 16 $Cu$
sites in the three-band Hubbard model and from
16 to 48 for the two other models.
For each model, the zero of the energy has been
set to the extrapolated value of the pair energy.}
\label{fig8}
\end{figure}

In Table 1, we list $t_{eff}$ normalized with respect to $\Delta_s$ 
for the $t-J$, and the one- and three-band Hubbard models with $J/t = 0.3$,
$U/t =$ 6 and 12, and $U_d/t_{pd} = 8$, $\Delta/t_{pd}=$ 2 and 3, respectively.
We have also
listed the pair binding energy $\Delta_{pb}$ in units of the spin gap
energy $\Delta_s$.
Now, as
previously discussed, the spin gap in the insulating case is of order $J/2$
and it sets the scale of the pair binding energy so that $\Delta_s$ and
$\Delta_{pb}$ for the three models are quite similar. However, the pair
dispersion is enhanced in going from the $t-J$ model to
the one-band Hubbard model and further enhanced for the three-band Hubbard 
model if one uses parameters appropriate for $CuO$ materials. 
We believe that this enhancement
is associated with the additional charge fluctuations which the 
one and three-band Hubbard models allow. 
The large enhancement in the three-band model reflects the fact that its 
charge gap is set by $\Delta$ when $U_d$ is large rather than $U_d$. 
In Fig.~9 we show the effective pair hopping as a function of $U/t$
in the one-band Hubbard model and as a function of $4t/J$ in the
$t-J$ model. 
In the $t-J$ model it seems that the effective pair hopping is always small
($t_{eff}/t < 0.4$).
In the one-band Hubbard model, however, the pair dispersion is strongly 
enhanced 
as $U/t$ decreases and for $U/t=6$ the ratio $t_{eff}/\Delta_s$ 
is similar to the value obtained in the three-band Hubbard model with 
$\Delta/t_{pd}=3$ and $U_d/t_{pd}=8$ (see Table~1).
For large values of $U/t$, the pair dispersion of the 
one-band Hubbard model approaches  that of the $t-J$ model
as shown in Fig.~9.
However, the difference between the pair dispersion in
both models seems to be of the order $4t/U$ when $U\rightarrow\infty$.
This difference could be due to next-nearest-neighbor
hopping terms of the order $4t/U$ which
are neglected in the derivation of the $t-J$ model
from the strong-coupling limit of the Hubbard model.

An effective magnon hopping $v_{eff}$ can be calculated from the  
dependence of the spin gap on the ladder length. We use
Eq.~\ref{eight} with $v_{eff}$ and $\Delta_s$ substituted for 
$t_{eff}$ and $\epsilon_p$, respectively. 
In this case the effective ladder length
is determined from the spin density profile.
In Table~1  $v_{eff}$ 
normalized with respect to $\Delta_s$
is listed for undoped ladders.

\begin{figure}[h]
\epsfxsize=3.375 in\centerline{\epsffile{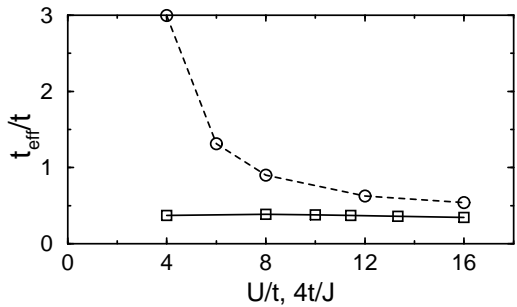}}
\caption{The effective hopping parameter for two holes is shown by
the circles for the one-band Hubbard model
and by the squares for the $t-J$ model.}
\label{fig9}
\end{figure}

\noindent The values of $v_{eff}/\Delta_s$ obtained in the three models 
are of the same order of magnitude for the physically relevant 
parameters.
We note that in this case the value obtained in the one-band
Hubbard model with $U/t=6$ is much larger than the other results.
In undoped ladders the effective magnon hopping obtained for 
the one-band Hubbard model approaches the 
result found in the $t-J$ (Heisenberg) model,
$v_{eff}\cong2.5 J$, at large values of $U/t$ as expected.
A recent perturbation calculation of two-leg Heisenberg ladders\cite{PS98}
gives $v_{eff}\cong2$, in satisfactory agreement with
our numerical result.


To summarize, all three models of course have a spin gap in the undoped phase.
Furthermore, the parameters for the different models can be chosen to make 
these spin gaps comparable and in the correct physical regime.
As noted, however, these three models have very different charge gaps
with the $t-J$ model having an infinite charge gap, the one-band Hubbard
model having a Mott-Hubbard gap set by $U$ and the three-band $CuO$
model having a charge transfer gap set by 
$\Delta= \varepsilon_p-\varepsilon_d$ 
for the physical parameter range of interest. 
The binding energy of two added holes is basically set by the spin gap.
However, $\Delta_{pb}/\Delta_{s} \cong$ 0.7 to 0.8 for
the $t-J$ and one-band Hubbard ladder 
while $\Delta_{pb}/\Delta_{s} \cong$ 1.2 to 1.6 for the three-band 
Hubbard ladder 
with parameters appropriate for $CuO$ materials.
Thus the hole pairs are in fact bound more tightly in units of the spin
gap energy in the three-band Hubbard ladder.
The main difference between the models is the size of the effective 
pair hopping which is significantly larger in the three-band model
for the physical parameters.
It is often assumed that these three models with parameters and band-fillings
appropriate for $CuO$ materials describe the same low-energy physics.
Our study has not revealed any fact which explicitly contradicts 
this point of view.
However, our results show that neither the one-band Hubbard model
nor the $t-J$ model can reproduce all the three-band Hubbard model 
results with a single set of effective parameters.
For instance, it seems that the pairing properties are better
reproduced by the one-band Hubbard model with $U/t=6$ than with the
usual parameters $U/t=12$.

Finally, we believe that the larger dispersion signified by the
effective pair hopping, which we found for the three-band Hubbard model,
points to an important feature of the physics contained in the charge
transfer insulator.
Pairs are less likely to localize in the three-band Hubbard model
than in the other models for the physical parameters.
This could have a significant effect in a two-dimensional lattice. 
For instance, in the two-dimensional $t-J$ model 
doped holes tend to form ordered arrays called stripes, which seem to
suppress superconductivity, at least when the stripes are static.\cite{WS97b}
A study of a three-leg ladder using both density matrix renormalization 
group methods and quantum Monte Carlo simulations
has shown that similar stripes also appear in the one-band Hubbard model
for $U/t>6$, but not for weaker coupling.\cite{Bonca} 
Thus, it seems that the sharp increase of pair mobility observed 
for decreasing $U/t$ in the two-leg ladder 
correlates with a transition from a striped ground state to 
a ground state without stripes in wider ladders.
Therefore, we think that the larger charge fluctuations which the three-band 
Hubbard model allows (compared to the $t-J$ model)
should lead in the two-dimensional lattice to a reduced tendency 
for domain walls to lock up in static arrays with suppressed 
superconductivity.

\acknowledgments
We thank R.L. Martin for helpful discussions. 
DJS acknowledges support from the Department of Energy under
grant DE-FG03-85ER-45197. DJS would also like to
acknowledge the Program on Correlated Electrons at the Center
for Material Science at Los Alamos National Laboratory.
SRW wishes to acknowledge the support of the Campus Laboratory
Collaborations Program of the University of California and
from the NSF under Grant No. DMR-9509945.
EJ thanks the  Swiss National Science Foundation for 
financial support.

\end{document}